\documentclass{extreport}
\usepackage[latin9]{inputenc}
\usepackage[a4paper]{geometry}
\usepackage{color}
\usepackage{array}
\usepackage{textcomp}
\usepackage{url}
\usepackage{amsmath}
\usepackage{amssymb}
\usepackage{graphicx}
\usepackage{setspace}
\usepackage[unicode=true,
 bookmarks=true,bookmarksnumbered=false,bookmarksopen=false,
 breaklinks=false,pdfborder={0 0 1},backref=section,colorlinks=true]
 {hyperref}
\hypersetup{pdftitle={Examplar-based Synthesis from Audio Features: Can One Resynthesize the Million Song Dataset?},
 pdfauthor={M.Moussallam, A. Liutkus, L.Daudet}}

\makeatletter

\newcommand{\noun}[1]{\textsc{#1}}
\providecommand{\tabularnewline}{\\}

\usepackage{cite}

\title{Listening to features}

\makeatother

\begin{document}
\begin{titlepage}
\newcommand{\HRule}{\rule{\linewidth}{0.5mm}} 
\center 
\textsc{\LARGE Technical Report}\\[1.5cm]
\HRule \\[0.4cm]
\textsc{\Large Institut Langevin}\\[1.5cm] 
\textsc{\Large ESPCI - CNRS - Paris Diderot University - UPMC}\\[0.5cm] 
\textsc{\large 1 rue Jussieu 75005 Paris France}\\[0.5cm] 

\HRule \\[0.1cm]
\textsc{Corresponding author: laurent.daudet@espci.fr}\\[0.5cm]

\author{Manuel Moussallam, Antoine Liutkus, Laurent Daudet}
\maketitle
\begin{abstract}
This work explores nonparametric methods which aim at synthesizing
audio from low-dimensionnal acoustic features typically used in MIR
frameworks. Several issues prevent this task to be straightforwardly
achieved. Such features are designed for analysis and not for synthesis,
thus favoring high-level description over easily inverted acoustic
representation. Whereas some previous studies already considered the
problem of synthesizing audio from features such as Mel-Frequency
Cepstral Coefficients, they mainly relied on the explicit formula
used to compute those features in order to inverse them. Here, we
instead adopt a simple blind approach, where arbitrary sets of features
can be used during synthesis and where reconstruction is exemplar-based.
After testing the approach on a speech synthesis from well known features
problem, we apply it to the more complex task of inverting songs from
the Million Song Dataset. What makes this task harder is twofold.
First, that features are irregularly spaced in the temporal domain
according to an onset-based segmentation. Second the exact method
used to compute these features is unknown, although the features for
new audio can be computed using their API as a black-box. In this
paper, we detail these difficulties and present a framework to nonetheless
attempting such synthesis by concatenating audio samples from a training
dataset, whose features have been computed beforehand. Samples are
selected at the segment level, in the feature space with a simple
nearest neighbor search. Additionnal constraints can then be defined
to enhance the synthesis pertinence. Preliminary experiments are presented
using RWC and GTZAN audio datasets to synthesize tracks from the Million
Song Dataset.
\end{abstract}
\end{titlepage}

\chapter{Introduction\label{sec:introduction}}

\section{From audio to features}

Audio features~\cite{mckinney2003features,mathieu2010yaafe} are
mid-level characteristics such as pitch, Mel-Frequency Cepstral Coefficients
(MFCC), loudness etc., which are computed from audio signals and whose
purpose is to serve as meaningful observations in audio machine learning
tasks. For example, they are fundamental in fingerprinting systems,
that consist in recognizing a whole musical song from a database,
based on the distorted measurement of an excerpt of a few seconds
only~\cite{wang2003fingerprint,miotto2008fingerprint,dupraz2010robustfingerprint,fenet2011scalablefingerprint}.
The use of features such as MFCC is also paramount in many classical
automatic speech transcription systems~\cite{rabiner1993fundamentals}
and, more generally, in most audio information retrieval studies.

\subsection{... and backwards ?}

Audio features are most commonly used in an \emph{analysis} setup,
where they are practical proxies that yield meaningful representations
for machine learning algorithms to work on. For \emph{synthesis} purposes,
only a limited number of them, such as MFCC along with pitch and loudness
information, are commonly used to control parametric speech synthesizers~\cite{chazan2000speechMFCC,milner2002speechMFCC,shao2004pitchMFCC,Ellis05-rastamat,milner2006cleanMFCC}.
However, these synthesis methods typically exploit the explicit knowledge
of how the features are computed in order to inverse them. In the
case of MFCC, for example, it is straightforward to relate the coefficients
back to the spectral envelope of the signal, thus permitting synthesis.

Now, suppose you have access to features that are obtained through
an unknown or complicated process and where no inverse operation that
permits to build back a sensible audio signal is available. Such features
may for example occur in a complicated fingerprinting system, whose
precise process is unknown, or could be provided as part of a dataset.
Lacking explicit inverse formulas, we can nonetheless assume to have
a \emph{development~database} at our disposal, which is composed
of both audio signals and their corresponding features. When observing
some \emph{test~features}, the proposed technique consists in mapping
the test features to the closest development features.

\section{Practical challenge: inverting the Million song dataset}

The million song dataset (MSD \cite{Bertin-Mahieux2011}) is a collection
of features collected over a very large number of audio tracks. On
top of metadata (or semantic features) such as Artist name, year of
publication, etc.. a few acoustic features are also provided for each
song. These features are provided by The Echo Nest%
\footnote{http://the.echonest.com%
} and can be obtained through the use of their API \cite{echonestAPI}.
In the FAQ section of the website%
\footnote{http://labrosa.ee.columbia.edu/millionsong/%
}, the answer to the question \emph{``Can I recover the audio from
the features?}'' is : ``\emph{Well.. you should try}''. This work
presents our first attempts to address this challenge. As implied,
this reverse engineering process is complex and one can list at least
three major issues:
\begin{itemize}
\item The acoustic features provided are rather scarce. On a typical audio
track, only a few dozens of them per second are available. 
\item The features are computed on non-overlapping slices of the audio data
called segments, whose boundaries are determined by an onset detection
routine. Therefore segments lengths may vary drastically.
\item The exact parameters of the feature calculus are not known, although
an extensive description of the analysis framework can be inferred
from Jehan's PhD dissertation \cite{Jehan2005}.
\end{itemize}
A first class of resynthesis methods typically use the acoustic features
as synthesis parameters. However, this requires that the features
are sufficiently adapted to the signal nature (e.g. speech reconstruction
from MFCC along with pitch and/or loudness information \cite{milner2002speechMFCC,chazan2000speechMFCC}).
For musical signals, in spite of existing attempts, the information
lost in the analysis process usually prevents any successful reconstruction.
Such synthesis from the MSD features has nonetheless been implemented
by Ellis%
\footnote{http://labrosa.ee.columbia.edu/millionsong/pages/matlab-introduction\#3%
}(see also \cite{EllisOnline}) and can serve as a baseline comparison.

The second class of methods perform synthesis by combining samples
from a training dataset. In this setup, the features are used in a
similarity search over the examples available. Among existing methods,
concatenative music synthesis systems rely on high fidelity features
as well as additional semantic information such as the musical score,
instrument and/or speaker (see the overview by Schwarz \cite{Schwarz2006}). 

In this work, we investigate nonparametric, exemplar-based synthesis
of audio tracks from an arbitrary set of acoustic features. This report
is organized as follows. The Nearest neighbor search is presented
in Chapter \ref{chap:Nearest-Neighbors-in}. Various Synthesis methods
are exposed in Chapter \ref{chap:Synthesis-methods}. Finally, Chapter
\ref{sec:Evaluation} presents our experiments. First in a controlled
environment where the features are computed by ouselves. Then we apply
it to the practical challenge of inverting the MSD. Although evaluation
of such system is uneasy, we provide some experiments in Section 5,
using RWC Music Genre \cite{Goto2003a} and GTZAN\cite{Tzanetakis2000}
Databases and a specifically designed objective reconstruction measure.
Audio examples are provided online.

\chapter{Nearest Neighbors in the feature space\label{chap:Nearest-Neighbors-in}}

\section{Framework}

\begin{figure*}[t]
\includegraphics[width=1\textwidth]{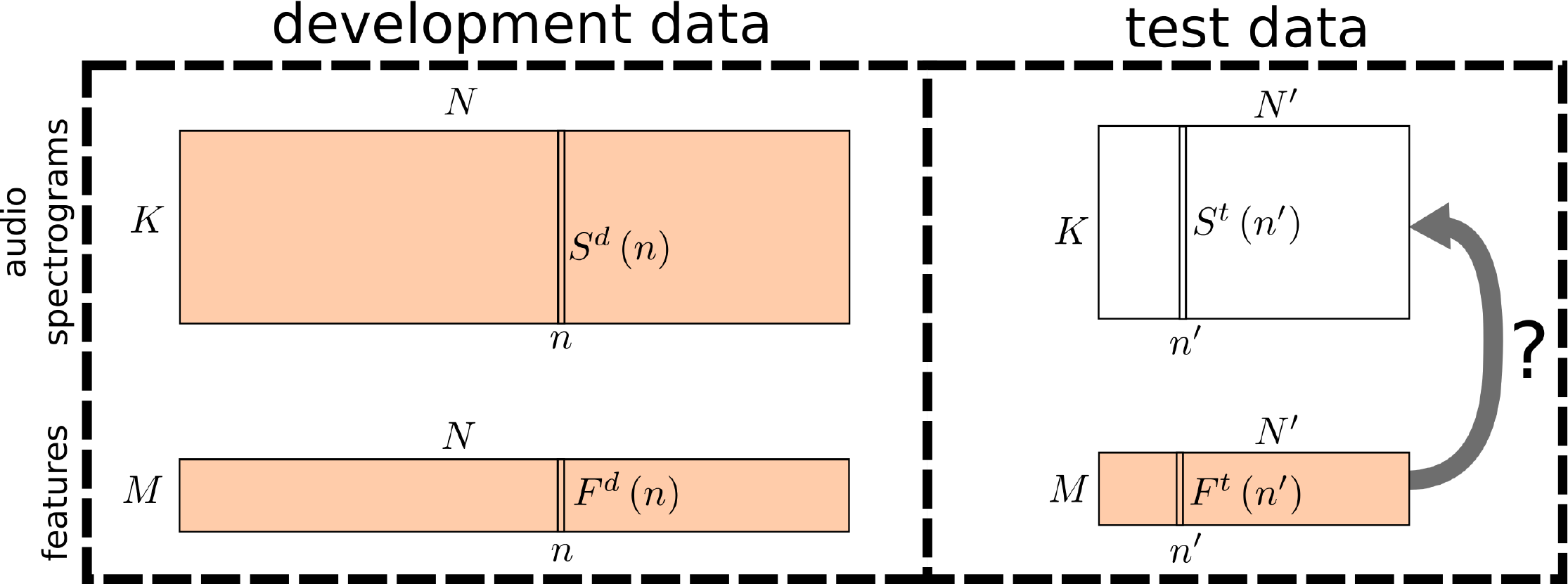}

\protect\caption{Main notations of the proposed setup. $S^{d}$ is a~$K\times N$
matrix gather the development-spectra, where $K$ and $N$ are the
number of frequency bins and frames, respectively. $F^{d}$, of dimension~$M\times N$,
provides the corresponding development features. During test, the
objective is to estimate audio spectrograms $S^{t}$, based on the
mere observation of the corresponding features~$F^{t}$.\label{fig:Main-notations}}
\end{figure*}

The proposed framework is summarized in Figure~\ref{fig:Main-notations}.
We assume that an audio \emph{development database }(dev-data) is
available, which typically consists of several hours of continuous
audio data. 

In any case, this dev-data can be analyzed through a Short Term Fourier
Transform (STFT), whose element-wise modulus is a $K\times N$ nonnegative
matrix $S^{d}$, where~$K$ and~$N$ respectively denote the number
of frequency bins and the number of frames. $S^{d}\left(n\right)$
and $F^{d}\left(n\right)$ will denote development data at frame~$n$,
understood as~$K\times1$ and $M\times1$ vectors, respectively.

The main idea of this study is to consider that \emph{other} observations
of features $F^{t}$, of dimension~$M\times N^{\prime}$ are available,
which have been computed from an \emph{unknown }underlying signal,
whose $K\times N^{\prime}$ \emph{unknown} magnitude spectrogram is
denoted~$S^{t}$. As highlighted in Figure~\ref{fig:Main-notations},
the main idea here is to use those test-features in conjunction with
the dev-data so as to yield a meaningful estimate for $S^{t}$.

\section{Estimation method}

The proposed blind synthesis method operates on a frame-by-frame basis.
For each given~$M\times1$ test frame~$F^{t}\left(n^{\prime}\right)$
of features, it estimates the corresponding $K\times1$ magnitude
spectrum~$S^{t}\left(n^{\prime}\right)$. 

For one given test frame~$n^{\prime}$, the chosen approach is to
identify the $P$ feature frames among the dev-data, which are the
most similar to~$F^{t}\left(n^{\prime}\right)$, and to estimate~$S^{t}\left(n^{\prime}\right)$
as the median of the $P$ corresponding development spectra. This
technique is reminiscent from recent works in audio source separation~\cite{RafiiREPETsim,Liutkus2012,Fitzgerald2012},
where magnitude spectrograms of background music are estimated as
median values of properly chosen spectra.

Formally, let $\mathcal{X}$ bet a set of $M$ development audio segments
$x_{m}$ for which both audio and the feature vectors $F(x_{m})$
are known. Let $s_{i}$ be a target segment for which only its feature
vector $F(s_{i})$ is available, we are interested in finding: 
\begin{equation}
\hat{s}_{i}=\arg\min_{x_{m}\in\mathcal{X}}\|F(s_{i})-F(x_{m})\|\text{\texttwosuperior}\label{eq:pb_inv_feat-1}
\end{equation}
let $d\left(f,f^{\prime}\right):\mathbb{R}^{M}\times\mathbb{R}^{M}\rightarrow\mathbb{R}^{+}$
be a known \emph{distance kernel}, which indicates the difference
between two feature vectors, both of dimension~$M\times1$. Many
possible choices for such kernels are possible, such as the simple
euclidean distance~:
\[
d\left(f,f^{\prime}\right)=\sqrt{\sum_{m=1}^{M}\left(f_{m}-f_{m}^{\prime}\right)^{2}}.
\]
This choice of kernel assumes that all the features equally contribute
to the metric which might not be the case. A weighted formulation
of such distance between two feature vectors would thus be:
\begin{equation}
d_{W}(F(s),F(s'))=\sqrt{\sum_{j=1}^{J}w_{j}\left(f_{j}(s)-f_{j}(s')\right)^{2}}\label{eq:weighted_kernel-1}
\end{equation}
where $W=[w_{j}]_{j=1..J}$ is a vector of weights to be determined.
With these notations, finding the $P$ nearest neighbors of a segment
$s$ amounts to finding the $P$ smallest elements in the set $D_{W}(s)=\{d_{W}\left(F(s),F(x_{m})\right)|x_{m}\in\mathcal{X}\}$.

Obviously, the features must be normalized. In the following, we will
assume that all features have been standardized according to:
\[
f_{j}(x_{k})\leftarrow\frac{f_{j}(x_{k})-\mu_{j}}{\sigma_{j}}
\]
where $\mu_{k}$ and $\sigma_{j}$ are respectively the sample mean
and standard deviation of the development data $\{f_{j}(x_{m})|x_{m}\in\mathcal{X}\}$.
For each target segment~$s$, we compute the~$N\times1$ vector~$D_{s}$,
whose~$m^{\mbox{th}}$ entry is given by~: 
\begin{equation}
D_{s}\left(m\right)=d\left(F^{t}\left(s\right),F^{t}\left(x_{m}\right)\right),\label{eq:distance_computation}
\end{equation}
and which basically gives the distance between current test feature
vector and the entries of the development database. Then, the indexes
of the~$P$ smallest elements of $D_{s}$ are identified, yielding
the indexes of the development frames which are most likely to be
similar to current test frame.

\chapter{\label{chap:Synthesis-methods}Synthesis methods}

\section{Concatenative Synthesis\label{sec:Concatenative-Synthesis-1}}

\subsection{Segment-by-segment approach}

\begin{figure}
\begin{centering}
\includegraphics[width=8cm]{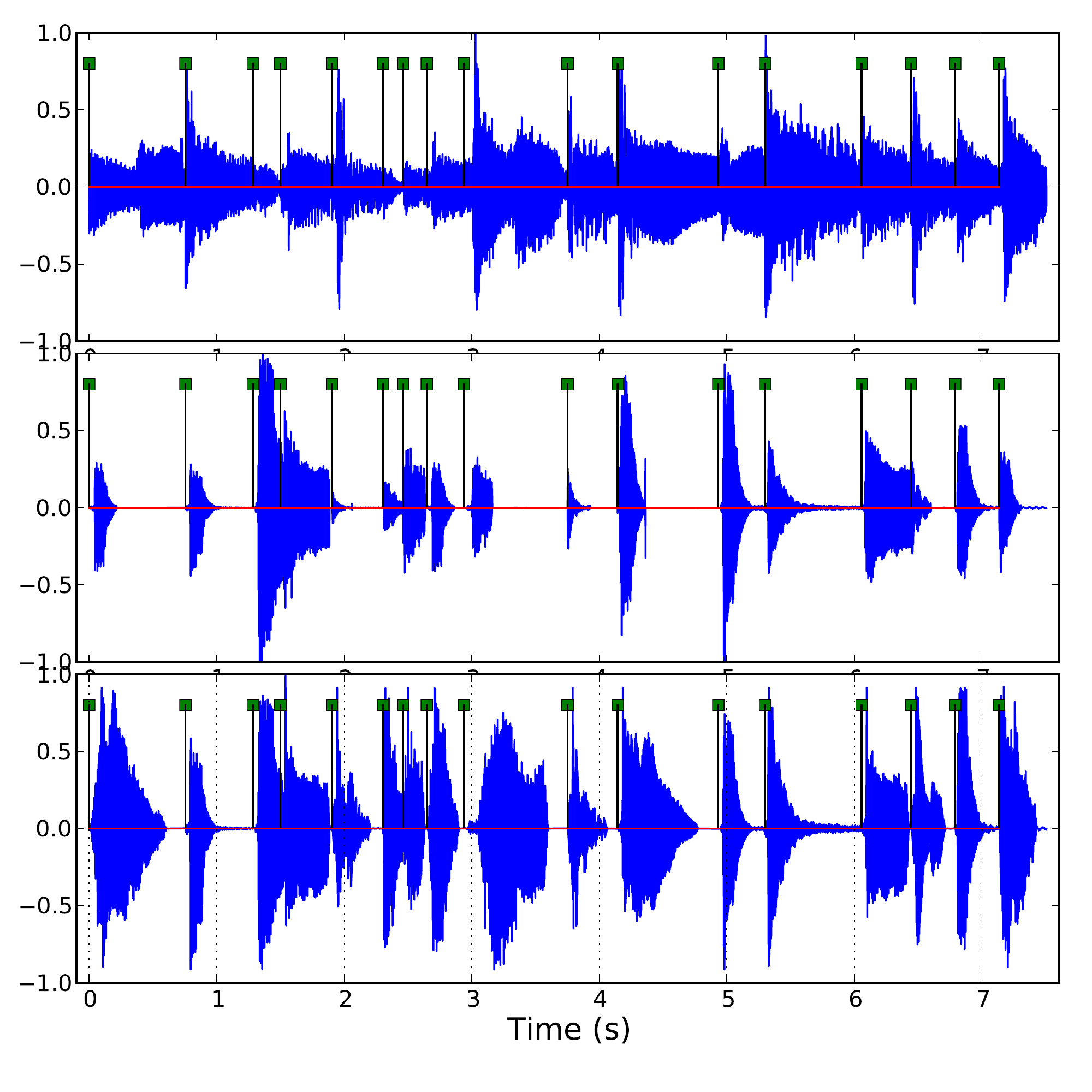}
\par\end{centering}

\protect\caption{Concatenative (cross) synthesis of a few seconds of blues from GTZAN
base using segments from the RWC Instrumental (Piano) database. The
mapping is performed using the Chroma vector computed through the
Echo Nest API. Upper: original waveform, Middle: synthesis using the
unaltered segments, Lower: synthesis using normalized and time-stretched
segments.\label{fig:Concatenative-(cross)-synthesis-1}}
\end{figure}

The simplest approach to synthesis is to work on each segment separately.
An estimate of a signal $x$, sliced into $I$ segments $s_{i}$ whose
positions $\{t_{i}\}_{i=1..I}$ are known, is obtained by replacing
each segment $s_{i}$ by the smallest element $\hat{s}_{i}$ of $D(s_{i})$,
relocated at $t_{i}$. This process yields an estimate $\hat{x}$
of $x$ defined by:
\begin{equation}
\hat{x}=[\hat{s}_{1}|\ldots|\hat{s}_{i}|\ldots|\hat{s}_{I}]
\end{equation}
where:
\[
\forall i\in[1..I],\hat{s}_{i}=\arg\min_{x_{m}\in\mathcal{X}}d_{W}\left(F(s_{i}),F(x_{m})\right)
\]
This simple scheme is sufficient to perform some kind of cross-synthesis
and we label this setup as the \emph{Cross-Plain} method. Figure \ref{fig:Concatenative-(cross)-synthesis-1}
shows an example of a target signal being synthesized using segments
from the RWC Instrumental Piano dataset. In this setup, the nearest
neighbor distance used relies only on the Chroma features (i.e. $w_{j}=1$
if $j\in[1..12]$, 0 otherwise).
 This simple example already raises some questions:
\begin{itemize}
\item How should the replacing segments $\hat{s}_{i}$ be normalized? 
\item Should the temporal alignment be somehow more relevant ? In particular,
the replacing segments and the original ones may have very different
length.
\end{itemize}
To address these issues, one may want to use the provided features
to control the synthesis. For instance, the replacing segment may
be transformed so that its resulting Loudness matches the target one
(provided by the first Timbre coefficient). A precise morphing could
additionally be controlled by the knowledge of the target Loudness
peak location. Lower plot in Figure \ref{fig:Concatenative-(cross)-synthesis-1}
shows an even simpler solution: segments are time-stretched to match
their target's length and normalization is achieved by ensuring all
segments waveform have same peak value. We label this setup as the
\emph{Cross-Normalized} method

As one could expect, these simple processing tricks are not sufficient
to reduce the main disturbing artifacts arising with such methods:
brutal transitions between consecutive segments. Tackling this issue
require either further processing of the transitions (e.g. fade-in/out)
or modifying the segment selection criterion to enforce some kind
of coherence between a chosen segment $\hat{s}_{i}$ and (at least)
its neighbors $\hat{s}_{i-1}$ and $\hat{s}_{i+1}$.

\subsection{Enforcing coherence, a regularization formulation}

More generally, one can think of many ways to modify the selection
according to various signal coherence criteria. The selection may
then be expressed as a penalized version of \eqref{eq:pb_inv_feat-1}:
\begin{equation}
\hat{s}_{i}=\arg\min_{x_{m}\in\mathcal{X}}d_{W}\left(F(s_{i}),F(x_{m})\right)+\lambda\mathcal{C}(x_{m},x)\label{eq:pb_inv_feat_penalized-1}
\end{equation}
where $\mathcal{C}$ can be any type of coherence cost between a sample
$x_{m}$ and the target signal $x$ (e.g. a stretching cost, a loudness
normalization cost, etc..). Expressing the concatenative synthesis
problem as a set of constraints is the approach adopted for instance
by Zils and Pachet in \cite{Zils2001}. In their work, they also define
\emph{sequence} constraints to enforce some kind of continuity between
the selected segments. In a similar manner, Schwarz \cite{Schwarz2000}
defines concatenative costs to assess the pertinence juxtaposing two
segments. 

A direct transposition of their methods to our setup is complex, since
both method relies on the availability of more detailed features (e.g.
pitch, spectral moments, etc..). Nonetheless, this pleads for longer-term
considerations to be considered. Abundant literature can be found
on the subject (interested reader may refer to the online survey maintained
at IRCAM%
\footnote{http://imtr.ircam.fr/imtr/Corpus-Based\_Sound\_Synthesis\_Survey%
}). An application to MSD-like features can be found in \cite{Jehan2005}. 

The nearest neighbor search yields for each target segment $s_{i}$
a set of $P$ candidates $\hat{s}_{i}^{p}$ with associated scores
$v_{i}^{p}=d_{W}(F(s_{i}),F(\hat{s}_{i}^{p}))$. Instead of choosing
the best candidate for each segment, one can search for an optimal
\emph{sequence} of candidates in the $P\times I$ grid. Given that
a transition cost between two segment $\mathcal{C}(\hat{s}_{i},\hat{s}_{i+1})$
is defined, a Viterbi algorithm can be used to identify this optimal
sequence. Possible choices for $\mathcal{C}(\hat{s}_{i},\hat{s}_{i+1})$
includes:
\begin{itemize}
\item The distance in the feature space $d_{W}\left(F(\hat{s}_{i}),F(\hat{s}_{i+1})\right)$
\item A fixed penalty cost $\lambda_{v}$ whose value depend on a priori
knowledge on segments similarities.
\end{itemize}
In this work, the audio datasets that we used are divided in audio
files, corresponding to parts of (e.g. 30 seconds in GTZAN), whole
songs (e.g. RWC Music Genre) or a chromatic scale played by a single
instrument (RWC Instrumental Piano). To favor coherence of selected
segments, we have investigated the following transition cost: 
\[
\mathcal{C}(\hat{s}_{i},\hat{s}_{i+1})=\begin{cases}
0 & \mbox{if }\hat{s}_{i}\mbox{ and}\hat{s}_{i+1}\mbox{ belong to the same file}\\
\lambda_{v} & \mbox{otherwise}
\end{cases}
\]
that favors the selection of segments belonging to the same audio
file. We label the concatenative synthesis using a Viterbi algorithm
and this transition cost as the \emph{Cross-Penalized }method.

\section{Additive Synthesis\label{sec:Additive-Synthesis-1}}

We have investigated a different approach, labeled additive synthesis.
Contrary to the concatenative synthesis where a single candidate per
segment is retained, additive synthesis will use a collection of examples
for each segment in order to build an estimate. Combining those elements
though, is not a trivial issue. A simple summation in the time domain
will give unsatisfactory results. The re-synthesized signal can be
modeled as a mixture of $P$ samples. Such mixture is better expressed
in the time-frequency domain. This technique is inspired by recent
works in audio source separation \cite{Fitzgerald2012,Liutkus2012}.

Let $s_{i}$ be a segment and $S_{i}$ be the modulus of its Short
Term Fourier Transform (STFT). The set of the $P$ smallest elements
in $D_{W}(s_{i})$ defines a set of $P$ segments $\{s_{i}^{p}\}_{p=1..P}$
with STFT modulus $\{S_{i}^{p}\}_{p=1..P}$. Since all these elements
are nonnegative, any positive combination of them remains nonnegative
and can therefore be considered as an estimated STFT modulus $\hat{S}_{i}$.
In this work, three types of combinations have been considered: 
\begin{align}
\hat{S}_{i}^{median} & =\mbox{median}_{p=1..P}\{S_{i}^{p}\}\\
\hat{S}_{i}^{mean} & =\frac{1}{P}\sum S_{i}^{p}\\
\hat{S}_{i}^{max} & =\max_{p=1..P}\{S_{i}^{p}\}
\end{align}
$\hat{S}_{i}^{median}$ was the only one being considered, for it's
ability to discard outliers. However, experiments showed that in some
situations, the opposite strategy of favoring outliers (i.e. $\hat{S}_{i}^{max}$)
gives interesting results. Finally, using $\hat{S}_{i}^{mean}$ realizes
a compromise between the two former strategies. Once a STFT modulus
$\hat{S}_{i}$ has been estimated, direct inversion is not possible
since the phase information is missing. In order to build a corresponding
audio waveform, some iterations of the classical Griffin and Lim algorithm\cite{Griffin1984}
can be used. We label the synthesis using (7), (8) and (9) respectively,
followed by a Griffin and Lim reconstruction as \emph{Add-Median},
\emph{Add-Mean} and \emph{Add-Max} methods.

\chapter{Evaluation\label{sec:Evaluation}}

\section{Speech synthesis from standard audio features}

\subsection{Experimental setup}

In this section, we propose an evaluation of the proposed method in
the context of speech synthesis. For this purpose, the dev-data considered
is a concatenation of~$10$s excerpts taken randomly from the \noun{Voxforge}
corpus%
\footnote{\url{www.voxforge.org}%
}., which consists of more than $3$ hours of speech signals uttered
both by male and female speakers and sampled at~$36$kHz. A STFT
is computed on the resulting waveform, with frames of~$32$ms (leading
to~$K=256$) and a hopsize of~$4$ms. $N$~depends on the size
of the development data and is one of the parameters of this evaluation. 

The test data is chosen as another excerpt from the \noun{Voxforge}
corpus, and corresponds to a sentence not found in the dev-data, being
uttered by a speaker also excluded from the dev-data. 

The same features were computed on both the development and test data
using the YAAFE~toolbox~\cite{mathieu2010yaafe}. Depending on~$M$,
common features were computed, as explained in Table~\ref{tab:features_table}.
As can be seen, the~$8$ first features were purposefully chosen
as highly non-invertible and the MFCC were chosen if $M$ gets higher,
due to their widespread use in speech processing. 

In this experiment, we perform synthesis using the \emph{Add-median}
method described in previous section.

\begin{table}
\begin{tabular}{|c|>{\raggedright}p{0.7\columnwidth}|}
\hline 
$M=3$ & \multicolumn{1}{>{\raggedright}p{0.7\columnwidth}|}{Zero crossing rate

Onset detection function 

Energy by frame}\tabularnewline
\hline 
\hline 
$M=8$ & Those above plus spectral slope, centroid, spread and flux \tabularnewline
\hline 
$M=11$ & Those above plus the~$3$ first Mel-Frequency Cepstral Coefficients
(MFCC)\tabularnewline
\hline 
$M=21$ & Those above plus $10$ more MFCC\tabularnewline
\hline 
\end{tabular}

\protect\caption{Features computed depending on~$M$\label{tab:features_table}.}
\end{table}

We report performance of the proposed method for varying sizes~$N$
of the development database, a varying number~$M$ of features and
a varying number~$P$ of neighbors selected for estimation. As an
objective metric, we consider the relative error~$20\log_{10}\frac{\left\Vert S^{t}-\hat{S^{t}}\right\Vert _{F}}{\left\Vert S^{t}\right\Vert _{F}}$
between original and estimated test spectrograms, where~$\left\Vert \cdot\right\Vert _{F}$
denotes the \noun{Frobenius} norm (root sum of squares). For each
choice of~$\left(M,N,P\right)$, $100$ independent tests were performed
and $E\left(M,N,P\right)$ is defined as the corresponding average
of the error obtained on all these tests. 

Even if this objective metric somewhat captures quality of reconstruction,
it is understood that only listening tests are fully relevant to evaluate
performance. To this purpose, a complete MATLAB implementation of
the proposed method, release under a BSD license, along with examples
of reconstructed signals are available on the webpage dedicated to
this paper%
\footnote{\url{www.example.com}%
}.

To the best of our knowledge, no other blind features inversion technique
similar to the one presented here has been presented so far and we
hence cannot compare its performance to previous comparable work.
Still, we chose to compare it nonetheless with an MFCC inversion technique~\cite{Ellis05-rastamat}
on the same data, which generates an audio signal based on a sequence
of MFCC, using explicit knowledge of the way they are computed, as
opposed to the blind inversion method described here.

\subsection{Results}

\begin{figure}
\begin{centering}
\includegraphics[width=8cm]{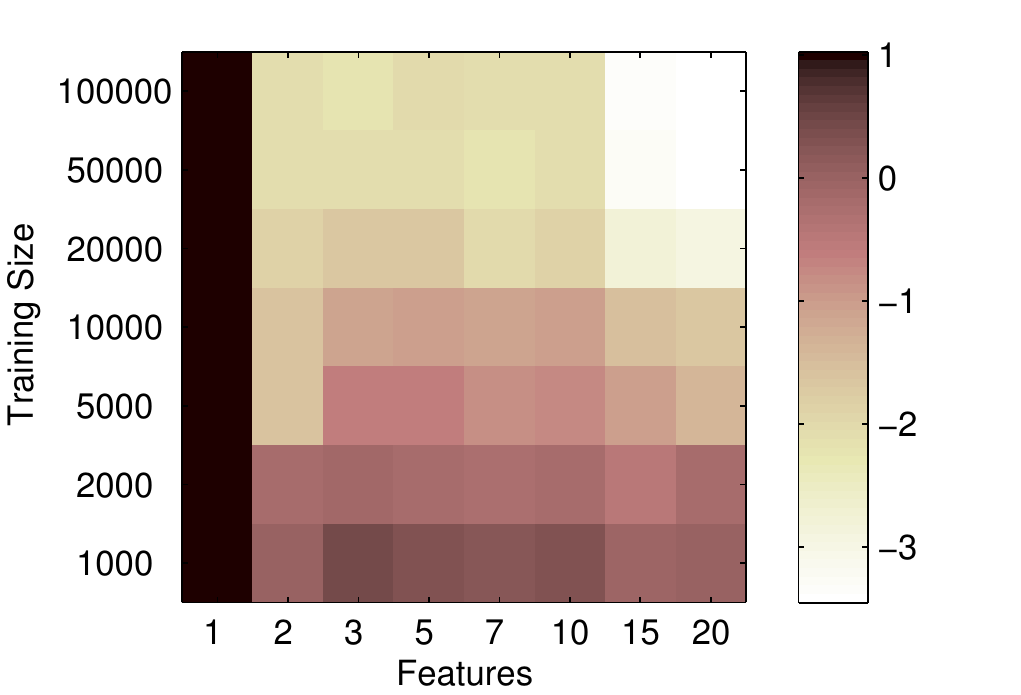}
\par\end{centering}

\protect\caption{Objective results: performance of the proposed method as a function
of the number of development frames~$N$ and number of features~$M$,
with~$P=10$ fixed. \label{fig:objective_results}}
\end{figure}

In figure~\ref{fig:objective_results}~(top), relative reconstruction
error~$E\left(M,N,P\right)$ is displayed as a function of $M$ and~$N$,
with~$P=10$ being fixed. As can be seen, increasing both~$M$ and~$N$
yields better results. Very interestingly, it can be seen that the
proposed method provides performance which is comparable to the deterministic
inversion of the MFCC, provided~$M$ and~$N$ are sufficiently large.
This result is very encouraging, since it means that blind inversion
is indeed a viable alternative to informed inversion approaches which
are dedicated to some specific features only.

In figure~\ref{fig:objective_results}~(bottom) is displayed~$E\left(M,N,P\right)$
as a function of~$P$ alone, with~$M=8$ and~$N=10^{5}$ being
fixed. As can be seen, the number of selected neighbors needs to be
high enough so as to smooth spurious matches, but also needs to be
small enough so as not to end up with an almost constant resulting
estimated spectrogram. However, performance of the method was found
to be rather robust to the choice of this parameter, and~$P=10$
generally seems like a good compromise.

\begin{figure}
\begin{centering}
\includegraphics[width=14cm]{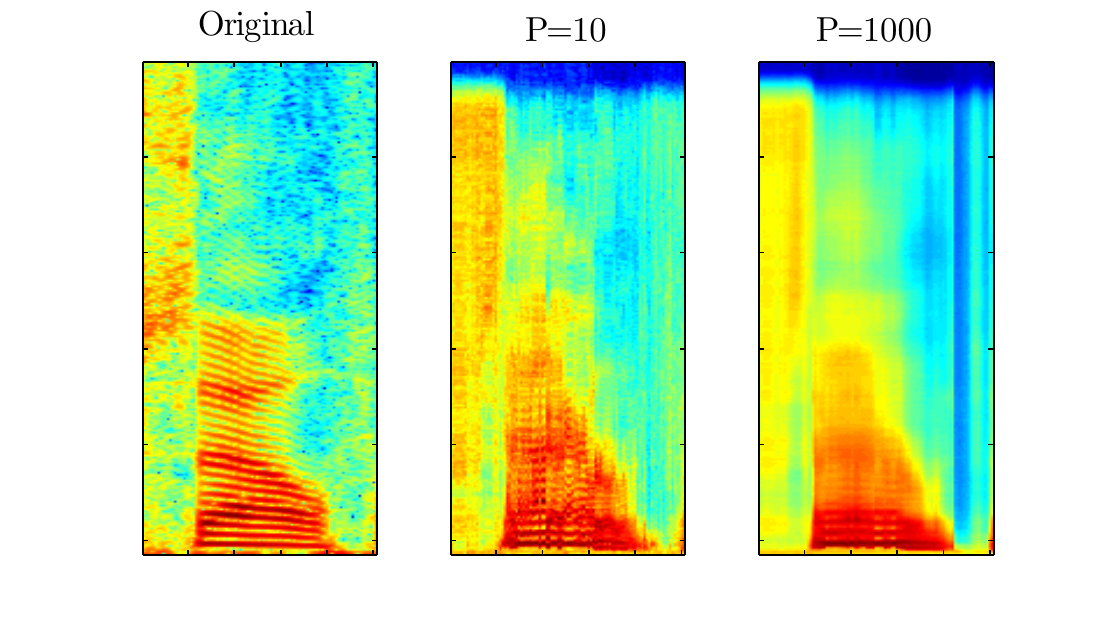}
\par\end{centering}

\protect\caption{Influence of $P$ on the reconstruction quality with~$M=7$ and~$N=10^{5}$
being fixed.\label{fig:objective_results-1}}
\end{figure}

\begin{figure}
\includegraphics[width=1\columnwidth]{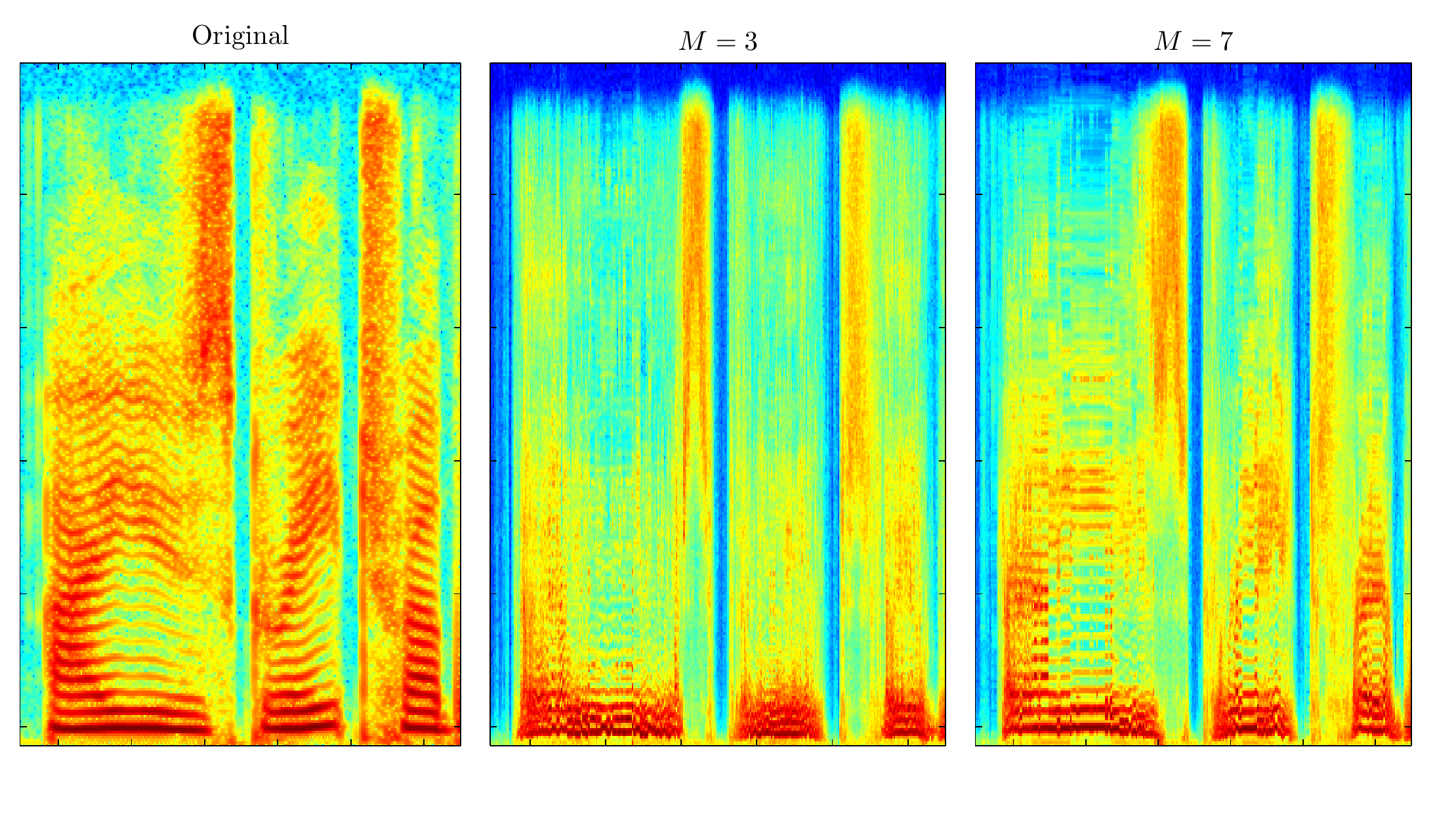}

\protect\caption{Typical reconstruction of a spectrogram using the proposed approach
with a development databased composed of~$N=10^{5}$ frames, corresponding
to~$6.6$min of continuous audio. Up~: original (unknown) test spectrogram
to be estimated. Middle~: estimate using $M=3$ features (zero-crossing
rate, onset detection function and frame energy). Down~: reconstruction
using~$M=8$ features ($M=3+$spectral centroid, slope, flux and
spread).\label{fig:spectrogram_reconstruction}}
\end{figure}

In figure~\ref{fig:spectrogram_reconstruction} is displayed one
particular example of reconstructed spectrogram using~$N=10^{5}$
development frames, $P=10$ and a varying number of features. As can
be seen, the estimated spectrogram is very similar to the original
one, using only highly non-invertible features.

\section{Inverting MSD songs}

\subsection{Overview of MSD features}

\begin{figure}
\begin{centering}
\includegraphics[width=8cm]{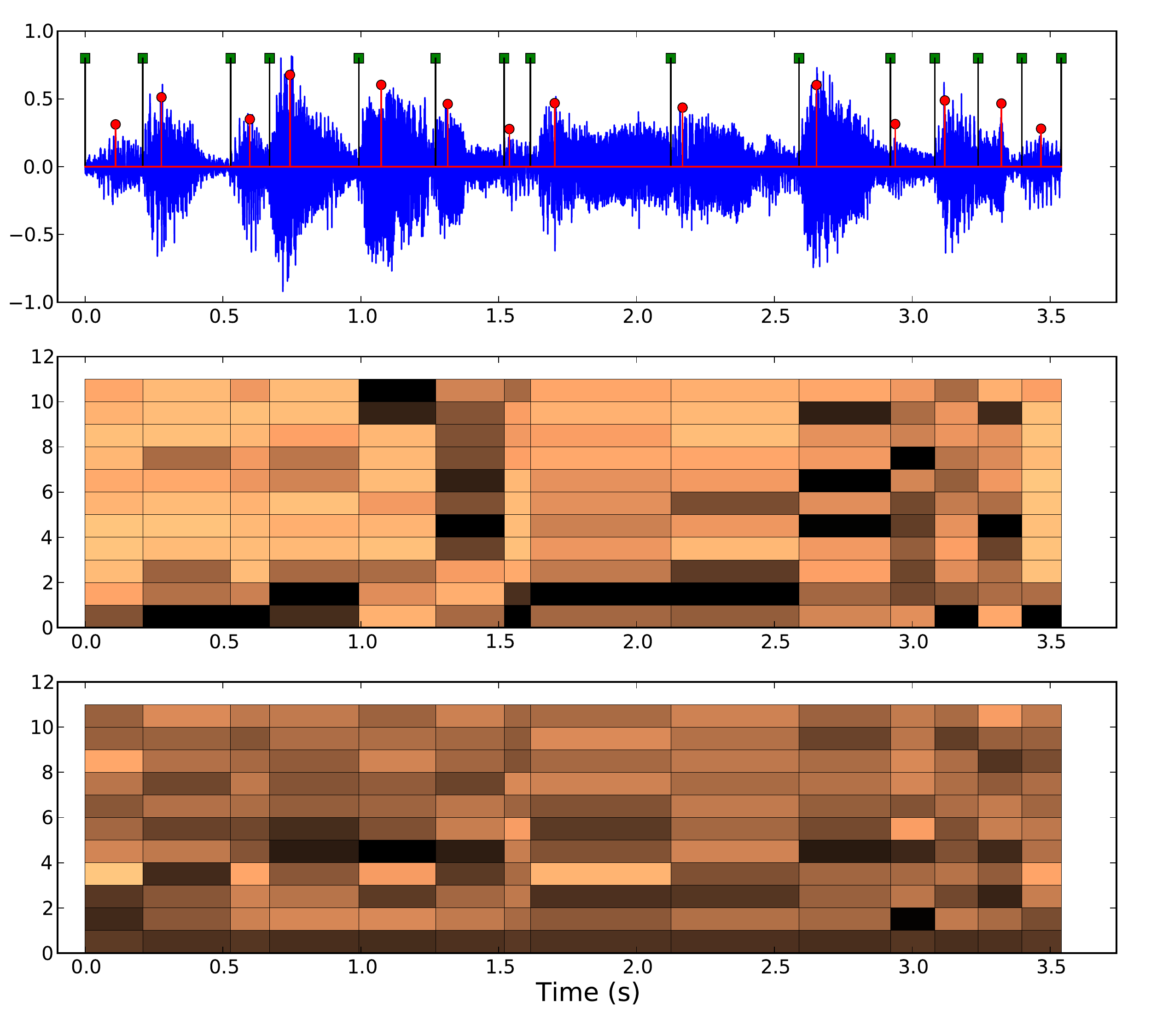}
\par\end{centering}

\protect\caption{Visualization of features fetched from the Echo Nest online API for
a short excerpt taken from the RWC database. Upper plot shows the
waveform. Segments boundaries are figured with green squares, loudness
peaks with red circles. The middle plot corresponds to the Chroma
features ($f_{1}$ to $f_{12}$) and the lower one to Timbre coefficients
($f_{13}$ to $f_{24}$) \label{fig:Visualization-of-features}}
\end{figure}

As explained on the MSD website, a signal $x$ is first sliced into
$I$ non overlapping segments:
\begin{eqnarray}
x & = & [s_{1}|\ldots|s_{i}|\ldots|s_{I}]
\end{eqnarray}
For each segment 27 acoustic features are computed (see \cite{Jehan2005,Ellis2010,Bertin-Mahieux2011}
for more details):
\begin{itemize}
\item 12 chroma coefficients $\{f_{j}\}_{j=1:12}$ describing the harmonic
content of the segment.
\item 12 timbre coefficients $\{f_{j}\}_{j=13:24}$ describing the sound
texture by quantifying its spectro-temporal shape. 
\item 3 Loudness coefficients : value at start $f_{25}$, value at peak
$f_{26}$ and peak position $f_{27}$. Loudness corresponds to the
perceived energy of the signal and relies on a bark scale nonlinear
mapping of the signal spectrum.
\end{itemize}
The feature vector $F(s)$ for a segment $s$ thus have the following
structure:
\begin{align}
F(s) & =[f_{1}(s)|\ldots|f_{j}(s)|\ldots|f_{J}(s)]
\end{align}
where $f_{j}(s)$ is the $j$-th feature of the segment $s$ and $J=27$.
In addition, the segmentation is provided in the form of the set $\{t_{i}\}_{i=1..I}$
of segment start instants. In average, there are about 4 segments
per second, but this may vary a lot depending on signal nature (e.g.
music tempo and rhythm). Figure \ref{fig:Visualization-of-features}
presents an overview of these features for a short audio excerpt.
In the given example, the total number of acoustic features available
is around 100 for one second of signal sampled at 44100 Hz. This drastic
dimensionality reduction is obviously a lossy process.

\subsection{Experimental setup}

As explained in \cite{Schwarz2006} (see also Chapter 6 in \cite{Jehan2005}),
results obtained through concatenative synthesis methods can generally
be considered a different piece of music. Therefore, measuring the
adequacy of the synthesized audio to the target is a challenging issue
that may require the use of a sound taxonomy, or carefully designed
listening tests.

Objective evaluation may yet be possible, when the original audio
is available, e.g. by using a distance in the time-frequency domain
between the original $S$ and an estimated STFT modulus $\hat{S}$.
A first idea would be to measure a mean-square error: 
\begin{equation}
MSE(S,\hat{S})=10\log_{10}\frac{\|S-\hat{S}\|_{F}}{\|S\|_{F}}
\end{equation}
where $\|.\|_{F}$ stands for the Frobenius norm or the sum of the
squares. For audio spectrograms, it is more relevant to measure the
Kullback-Leibler (KL) divergence of the normalized magnitude spectrograms,
seen as probability density functions:
\begin{equation}
KL(S_{N},\hat{S}_{N})=\sum_{n=1}^{L}\sum_{k=1}^{K}S_{N}(k,n)\log\left(\frac{S_{N}(k,n)}{\hat{S}_{N}(k,n)}\right)\label{eq:KLspec}
\end{equation}
where $n$ and $k$ are respectively the discrete time and frequency
index of the spectrograms, and normalization is achieved by: 
\begin{equation}
S_{N}(k,n)=\frac{S(k,n)}{\sum_{n}\sum_{k}S(k,n)}\label{eq:norm_spectro}
\end{equation}

We have used the provided API to gather features for the RWC Music
Genre (RWC-MDB-G-2001-M01$\sim$M09, 100 files), Instrumental Piano
(RWC-MDB-I-2001 n1$\sim$12, 12 files) and GTZAN (1000 files) datasets.
Combined, we have a collection $\mathcal{X}$ of $M_{max}\simeq170000$
segments corresponding to approximately 20 hours of audio data. Since
we need the real audio data for objective performance measurements,
we use part of this database for the test. Nonetheless, any of the
synthesis method can straightforwardly be applied to a file in the
MSD format (see the companion website for examples).

A development set $\mathcal{X}_{dev}\subset\mathcal{X}$ of $M$ segments
is first drawn by selecting files at random in the complete collection.
And the synthesis is evaluated on a 20-segment length excerpt $x_{test}$
chosen in a random file from the complementary set $\mathcal{X}_{test}=\overline{\mathcal{X}}_{dev}$.
To ensure that no exactly similar segments are simultaneously present
in the train and test sets (due to some redundancies in the GTZAN
dataset), we remove all segments belonging to the same genre as $x_{test}$.
We investigate the following parameters:
\begin{itemize}
\item Synthesis method: 6 of them are considered: 3 concatenative ones (\emph{Cross-Plain},
\emph{Cross-Normalized} and \emph{Cross-Penalized}, 3 additive ones
(\emph{Add-Mean}, \emph{Add-Max} and \emph{Add-Median}) and a parametric
one for comparison obtained using the software provided by D. Ellis%
\footnote{http://labrosa.ee.columbia.edu/millionsong/pages/matlab-introduction\#3%
}.
\item Combination of features used in the nearest neighbor search.
\item Number of neighbors considered $P$.
\end{itemize}

\subsection{Results}

\subsubsection{Audio examples}

The concatenative synthesis methods \emph{Cross-Plain, Cross-Normalized
}and\emph{ Cross-Penalized} perform erratically relative to the spectrogram
reconstruction metric \eqref{eq:KLspec}. We provide as many sound
examples as possible (both from our own dataset and the MSD) of the
synthesis results on a companion website%
\footnote{Anonymized version for review purposes: https://sites.google.com/site/tempsubismir13/%
}. Depending on the chosen method, very different types of results
can be obtained. This diversity of synthesis is better appreciated
by listening to examples. Nonetheless, some interesting observations
can be made with objective measurement of additive synthesis experiments.

\subsubsection{Influence of the feature combination}

The size of the development set is fixed to $M=100000$. We investigate
the 7 possible combinations of the three types of features (Chroma,
Timbre, Loudness). For each feature combination, 100 tests are run,
25 for each value of $P$ among $\{1,5,10,20\}$.

\begin{figure}
\begin{centering}
\includegraphics[width=8cm]{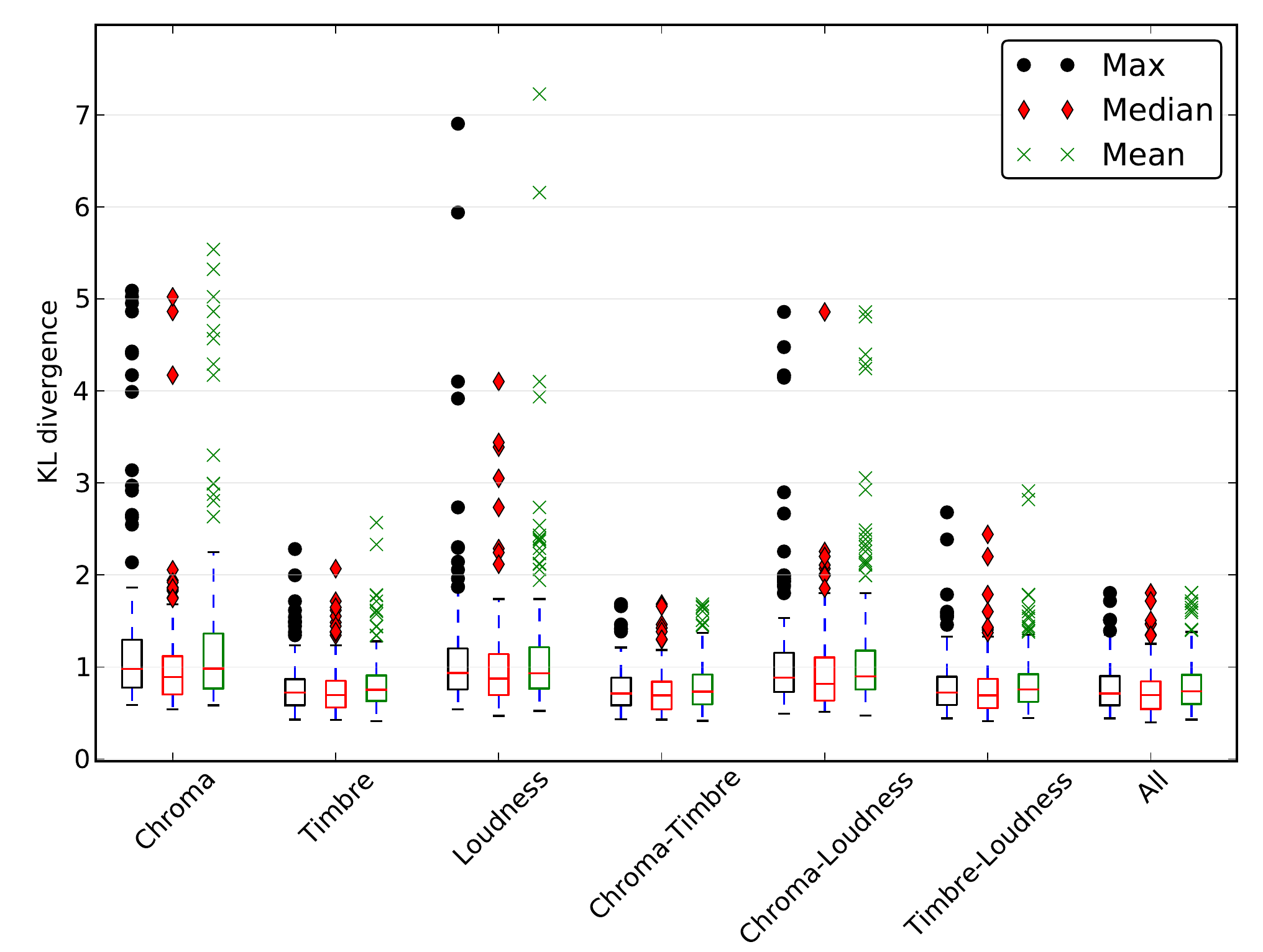}
\par\end{centering}

\protect\caption{KL divergence between normalized spectrogram of original and reconstruction
with (7), (8) and (9), results for 100 tests for each of the 7 features
combinations. Timbre features seem to be the most effective ones.\label{fig:Influence-of-feat_comb}}
\end{figure}

Figure \ref{fig:Influence-of-feat_comb} shows the normalized spectrogram
KL divergence between original and resynthesized results for the different
combinations of features and STFT modulus combining strategies. Timbre
coefficients seem more robust to the task than Chroma and Loudness
ones. Best results are arguably reached using Timbre+Chroma features,
and using Loudness features does not seem to improve the results in
any case. This experiments also shows that there seem not to be a
significant difference between the STFT modulus combining strategies,
although the synthesis may sound quite different. 

\subsubsection{Influence of $P$}

\begin{figure}
\begin{centering}
\includegraphics[width=8cm]{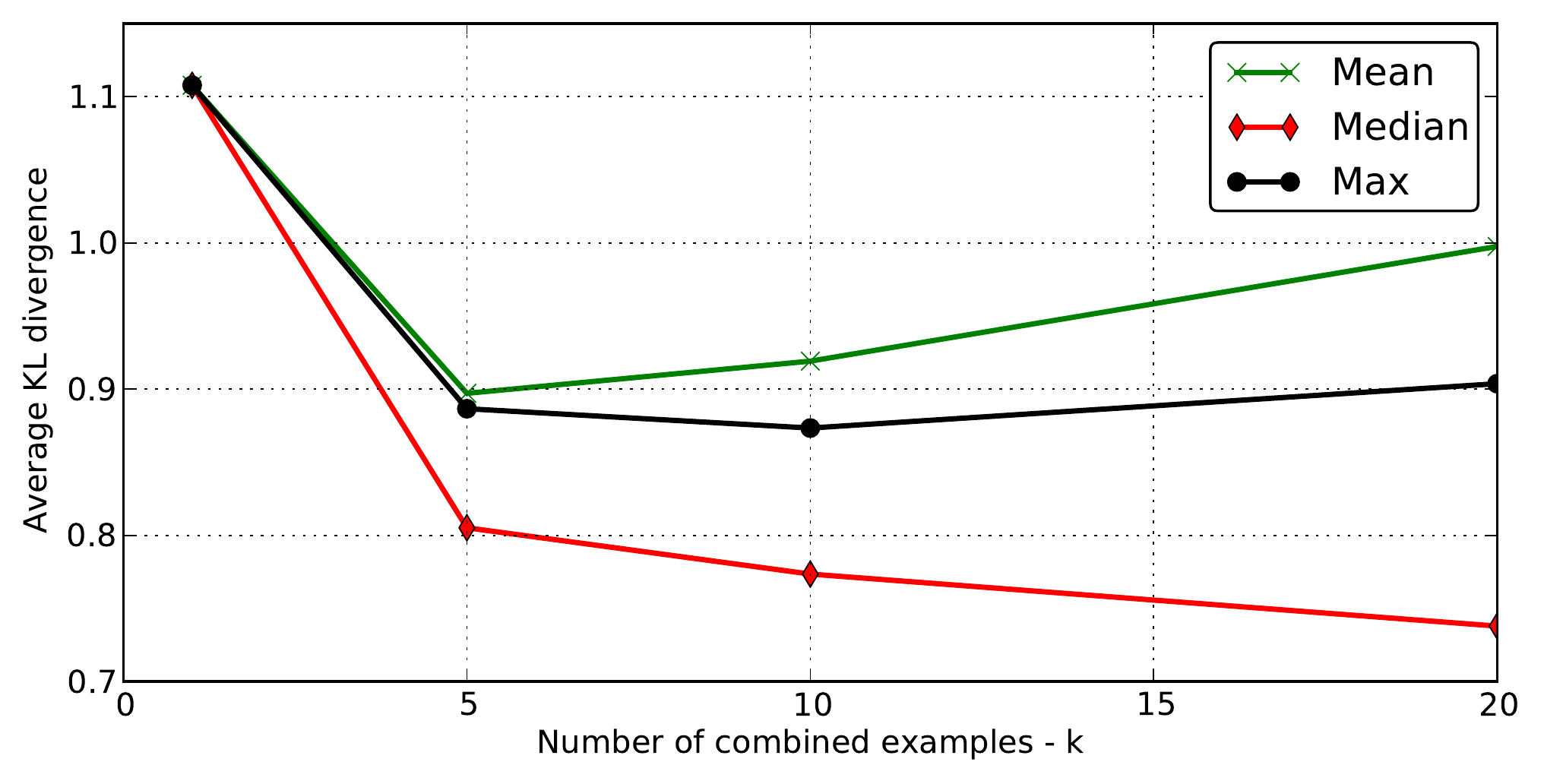}
\par\end{centering}

\protect\caption{KL divergence of reconstructed STFT modulus for various number of
considered examples\label{fig:Influence-of-K}.}

\end{figure}

With the same experimental setup, we can observe results marginalized
over $P$. Figure \ref{fig:Influence-of-K} summarizes the observations.
The fact that Mean and Max strategies seem to perform slightly worse
when $P>5$ is mainly due to the presence of outliers. This corresponds
to situations where one or more of the candidate is poorly correlated
but still takes over the other ones, this scenario being unlikely
to appear in the case of the median.

\begin{figure*}[t]
\begin{centering}
\includegraphics[width=16cm,height=18cm]{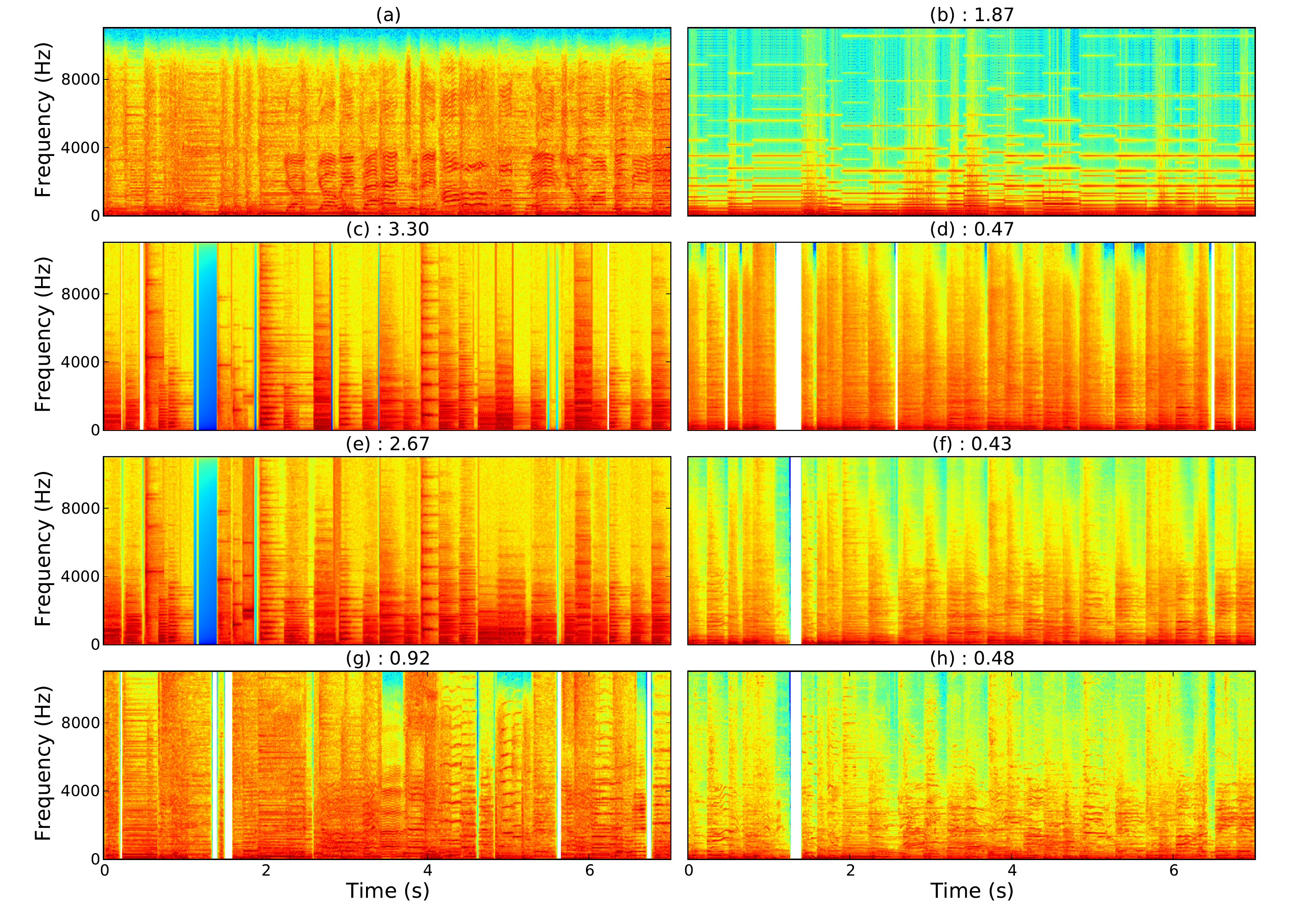}
\par\end{centering}

\protect\caption{Logarithm of the STFT modulus of (a) original, (b) parametric synthesis
using D. Ellis Matlab code. (c) \emph{Cross-Plain }and\emph{ (e) Cross-Normalized}
(using only RWC Instrumental Piano for $\mathcal{X}_{dev}$), (g)
\emph{Cross-Penalized} (using all dataset), (d) \emph{Add-Median,
(f) Add-Mean} and \emph{(h) Add-Max} for $P=10$. All nearest neighbor
search used the Chroma-Timbre combination except (c) and (e) for which
only the Chroma were used. Displayed values are the corresponding
KL divergence scores \eqref{eq:KLspec}.\label{fig:Log-spectrograms}}

\end{figure*}

Finally, Figure \ref{fig:Log-spectrograms} present the spectrograms
for an example for the 6 proposed strategies and the baseline parametric
synthesis. The corresponding sounds as well as many more examples
are available on the companion website.

\chapter*{Conclusion}

In this work, we present some early attempts to resynthesize audio
from scarce, non-uniformly distributed, arbitrary features as one
can found in the MSD. Synthesis is achieved through the use of a development
dataset for which the audio is available. After applying state of
the art approaches of concatenative synthesis to this context, we
propose a different scheme of additive synthesis that seems to help
achieve a pertinent reconstruction of the magnitude spectrogram.

Many improvements can be thought of. First of all, more data may be
needed. Raising the development dataset from 20 to 200 or even 2000
hours of audio would certainly improve the quality of the reconstruction.
Second, our methods are based on a purely non-supervised neighbor
search algorithm, obviously one could try supervised or semi-supervised
methods. Additionally, the weights of the distance kernel in \ref{eq:weighted_kernel-1}
could be learned to improve the pertinence of the selection. 

Future work will investigate these points. As one may notice, no use
is made in this work of all the metadata available in the MSD, nor
of other types of features such as tempo, beats etc.. Arguably the
reconstruction could benefit from this additional information. Finally,
other types of signal coherence constraints might be expressed as
regularizations of problem \eqref{eq:pb_inv_feat_penalized-1} (e.g.
sparsity -or structured sparsity - of the reconstructed spectrogram
in a given dictionary). MSD resynthesis remains a highly challenging
issue, but not an impossible one.

\bibliographystyle{ieeetr}
\bibliography{biblio}

\begin{thebibliography}{10}

\bibitem{mckinney2003features}
M.~McKinney and J.~Breebaart, ``Features for audio and music classification,''
  in {\em Proceedings of the International Society for Music Information
  Retrieval Conference (ISMIR)}, vol.~3, pp.~151--158, 2003.

\bibitem{mathieu2010yaafe}
B.~Mathieu, S.~Essid, T.~Fillon, J.~Prado, and G.~Richard, ``{Yaafe}, an easy
  to use and efficient audio feature extraction software,'' in {\em Proceedings
  of the 11th International Society for Music Information Retrieval Conference
  (ISMIR)}, 2010.

\bibitem{wang2003fingerprint}
C.~Wang and L.~Avery, ``An industrial strength audio search algorithm,'' in
  {\em Proceedings of the International Society for Music Information Retrieval
  Conference (ISMIR)}, vol.~3, 2003.

\bibitem{miotto2008fingerprint}
R.~Miotto and N.~Orio, ``A music identification system based on chroma indexing
  and statistical modeling,'' in {\em Proceedings of the International
  Conference on Music Information Retrieval}, pp.~301--306, 2008.

\bibitem{dupraz2010robustfingerprint}
E.~Dupraz and G.~Richard, ``Robust frequency-based audio fingerprinting,'' in
  {\em IEEE International Conference on Acoustics, Speech and Signal Processing
  (ICASSP)}, pp.~281--284, IEEE, 2010.

\bibitem{fenet2011scalablefingerprint}
S.~Fenet, G.~Richard, and Y.~Grenier, ``A scalable audio fingerprint method
  with robustness to pitch-shifting,'' in {\em Proceedings of the 12th
  International Conference on Music Information Retrieval (ISMIR)}, 2011.

\bibitem{rabiner1993fundamentals}
L.~Rabiner and B.~Juang, ``Fundamentals of speech recognition,'' 1993.

\bibitem{chazan2000speechMFCC}
D.~Chazan, R.~Hoory, G.~Cohen, and M.~Zibulski, ``Speech reconstruction from
  mel frequency cepstral coefficients and pitch frequency,'' in {\em IEEE
  International Conference on Acoustics, Speech and Signal Processing
  (ICASSP)}, vol.~3, pp.~1299--1302, IEEE, 2000.

\bibitem{milner2002speechMFCC}
B.~Milner and X.~Shao, ``Speech reconstruction from mel-frequency cepstral
  coefficients using a source-filter model,'' in {\em International Conference
  on Spoken Language Processing (ICSLP)}, pp.~2421--2424, Citeseer, 2002.

\bibitem{shao2004pitchMFCC}
X.~Shao and B.~Milner, ``Pitch prediction from mfcc vectors for speech
  reconstruction,'' in {\em IEEE International Conference on Acoustics, Speech
  and Signal Processing (ICASSP)}, vol.~1, pp.~I--97, IEEE, 2004.

\bibitem{Ellis05-rastamat}
D.~Ellis, ``{PLP} and {RASTA} (and {MFCC}, and inversion) in {M}atlab,'' 2005.
\newblock online web resource.

\bibitem{milner2006cleanMFCC}
B.~Milner and X.~Shao, ``Clean speech reconstruction from mfcc vectors and
  fundamental frequency using an integrated front-end,'' {\em Speech
  Communication}, vol.~48, no.~6, pp.~697--715, 2006.

\bibitem{Bertin-Mahieux2011}
T.~Bertin-Mahieux, D.~Ellis, B.~Whitman, and P.~Lamere, ``{The million song
  dataset},'' in {\em International Society for Music Information Retrieval
  Conference}, 2011.

\bibitem{echonestAPI}
``{The Echo Nest Analize, API},'' in {\em http://developer.echonest.com}.

\bibitem{Jehan2005}
T.~Jehan, {\em {Creating music by listening}}.
\newblock PhD thesis, Massachussets Institute of Technology, 2005.

\bibitem{EllisOnline}
D.~Ellis, ``{PLP and RASTA (and MFCC, and Inversion) in Matlab},'' in {\em
  http://www.ee.columbia.edu/\~{}dpwe/resources/matlab/rastamat/}, 2005.

\bibitem{Schwarz2006}
D.~Schwarz, ``{Concatenative sound synthesis: The early years},'' {\em Journal
  of New Music Research}, vol.~35, pp.~3--22, Mar. 2006.

\bibitem{Goto2003a}
M.~Goto and H.~Hashiguchi, ``{RWC music database: Music genre database and
  musical instrument sound database},'' in {\em Proc. ISMIR}, no.~October,
  pp.~229--230, 2003.

\bibitem{Tzanetakis2000}
G.~Tzanetakis and F.~Cook, ``{Sound analysis using MPEG compressed audio},'' in
  {\em IEEE International Conference on Acoustics, Speech and Signal
  Processing}, vol.~2, pp.~761--764, 2000.

\bibitem{RafiiREPETsim}
Z.~Rafii and B.~Pardo, ``Music/voice separation using the similarity matrix,''
  in {\em Proceedings of the 13th International Conference on Music Information
  Retrieval (ISMIR)}, pp.~583--588, 2012.

\bibitem{Liutkus2012}
A.~Liutkus, Z.~Rafii, R.~Badeau, B.~Pardo, and G.~Richard, ``{Adaptive
  filtering for music/voice separation exploiting the repeating musical
  structure},'' in {\em IEEE International Conference on Acoustics, Speech and
  Signal Processing}, pp.~53--56, 2012.

\bibitem{Fitzgerald2012}
D.~Fitzgerald, ``{Vocal Separation Using Nearest Neighbours and Median
  Filtering},'' in {\em Irish Signal and Systems Conference}, pp.~583--588,
  2012.

\bibitem{Zils2001}
A.~Zils and F.~Pachet, ``{Musical mosaicing},'' in {\em Digital Audio Effects
  (DAFx)}, pp.~1--6, 2001.

\bibitem{Schwarz2000}
D.~Schwarz, ``{A system for data-driven concatenative sound synthesis},'' {\em
  Digital Audio Effects (DAFx)}, pp.~1--6, 2000.

\bibitem{Griffin1984}
D.~Griffin and J.~Lim, ``{Signal estimation from modified short-time Fourier
  transform},'' {\em IEEE Transactions on Acoustics, Speech and Signal
  Processing}, vol.~32, no.~2, pp.~236--243, 1984.

\bibitem{Ellis2010}
D.~Ellis, B.~Whitman, T.~Jehan, and P.~Lamere, ``{The echo nest musical
  fingerprint},'' in {\em International Society for Music Information Retrieval
  Conference}, vol.~32, 2010.

\end{thebibliography}

\end{document}